\documentstyle[12pt]{article}
     \begin{document}
      \vskip 2. true cm
      \begin{center}
    {\large \bf Decoherence within a simple Model for the Environment} \\

      \vskip 1.4 true cm
        B. Carazza \\
       \vskip .1 true cm
\par 
       {\it Dipartimento di Fisica dell' Universit\`a,} \\
      {\it viale delle Scienze, I43100 Parma, Italy}\\
        {\it INFN Sezione di Cagliari, Italy}\\  
      \end{center}
\par \vskip 2mm
\begin{center} {\bf  Abstract} \end{center}
 \vskip 1mm
\begin{quote}
 This article examines the decoherence of a macroscopic body using a simple
 model of the environment and following the evolution of the pure state for
 the whole system. We found that decoherence occurs for very general initial
 conditions and were able to confirm a number of widely accepted
 features of the process.
\end{quote}
\vskip 1.5 true cm
Key words: Decoherence, reduced density matrix,
preferred basis, macrosuperpositions.
 \newpage
      \noindent{\bf 1. INTRODUCTION}
      \vskip .6 true cm
 \par 
 There is growing interest in decoherence, particularly as 
  it promises to deepen our
 understanding of the macroscopic world in terms of quantum principles
 \cite{omn}. If we suppose that $B$ is a macroscopic body
  (say a point particle
 with a very large mass M and position co-ordinate at the
 centre of mass) interacting with an environment $E$, we know from a
 host of examples that if we start with a factorized state of the two
  systems,
  the environment state vectors associated with the 
 body's distinct positions (at least
 macroscopically) rapidly become orthogonal
 to each other, irrespective of the initial conditions. 
 The spatial
 correlations of $B$, whose reduced density matrix becomes diagonal on a
 position basis, are suppressed by decoherence. Thanks to this  apparent
 collapse  due to interaction with the environment, we see the classic
 characteristics of the macroscopic world emerge.
 During the above mentioned process temporal evolution also affects body
 $B$. In a number of examples, however (some in a seminal paper by Joos
 and Zeh\cite{zeh}), wave packet spreading is ignored.
 This is justified both by the speed of the decoherence process 
 and because mass $M$ is large. We will adopt this approximation 
   by formally ignoring the term of the body's kinetic
 energy in the total Hamiltonian. This is rigorous with the limiting
 condition  that $M$ is infinite.
 Realistic sources of decoherence are described in the literature
 on the subject, such as scattering\cite{zeh,gal,teg} 
 and quantum gravity\cite{hav,ell} but the focus is
 generally on useful and practical models such as the so-called Caldeira-
 Legett environment \cite{cal}, which in any case reflects certain physical
 situations. Omnes \cite{omn2} recently put
  forward a general theory of the decoherence effect 
 encompassing both Caldeira-Legett's harmonic model and
 the external environment considered by Joos and Zeh\cite{zeh}, 
  as well as being related to
 the quantum state diffusion model\cite{qua}.
 In the applications mentioned, the usual approach is to study the
 temporal evolution of the reduced density matrix we are interested in,
 which obeys an irreversible master equation\cite{joo}.
  Resorting to this
 technique is almost inevitable as the overall system has many 
 (or infinite) degrees of freedom. However, in my opinion at least,
 it would be useful to propose a further example of
 decoherence within the framework of an environment model using
 Schroedinger evolution of the initial pure state.
 Since the seminal paper by H. D. Zeh, published in the very
 first issue of Foundations of Physics\cite{zzz}, our understanding
 of the decoherence process has gradually shifted from qualitative
 to quantitative. Our aim is to use pure state evolution
 to obtain analytical results independent
  of a particular master equation.
 While there will be no 
 ground breaking results, a new example
 could provide a further confirmation of our convictions, 
 as well as offering a number of 
 points worthy of consideration. This means the paper is didactic
 to a certain degree, but also represents a small step forward in our
 understanding of the decoherence process in quantitative terms.
 
 \vskip .9 true cm
 \noindent{\bf 2. THE MODEL FOR THE ENVIRONMENT }
 \vskip .6 true cm
 \par
 The example we will discuss concerns a one dimensional system
 in which the external environment is represented by $n$ identical
 particles with $x$ axis co- ordinates $x_1, x_2, \ldots, x_n$.
  The macroscopic body's position
 is indicated by $X$. The Hamiltonian for the environment is based on the
 Hepp-Coleman\cite{hepp} model, which appeared some time ago and was also
 discussed by Bell\cite{bel}.
 It was used to discuss the problem of measurement in terms
 of  apparent collapse. (It was actually adopted to describe the
 system to observe, whereas here it will be used to describe
 the environment which acts as observer). The model considers the
 particle's kinetic energy as proportional to its momentum
 rather than as given by the customary quadratic term. The wave packet
 of a free particle of this sort moves along the axis in a single direction
 without changing form. 
 Our Hamiltonian for the environment is
       \begin{equation}\label{e1}
       H_e= \alpha \sum \hat p_i + V(x_1,x_2, \ldots, x_n)
       \end{equation}
 where $\hat p_i$ is the momentum,  $\alpha$ is a constant quantity
 with the dimension of a velocity, and the potential
 term $V(x_1,x_2, \ldots)$ includes both
 one particle interactions and two or many particle interactions.
  The expression assumed for the kinetic term is rather 
  unconventional and the hamiltonian is considered to be 
  an appropriately idealized toy model. However, the intuition
  can be supported by reference to physical models.
   In the case of a free particle, for example, it may be considered
    as the caricature of an electron in one dimensional motion. 
   Let us consider the ultra-relativistic  approximation frequently
  used for very energetic electrons, and
   which amount to take the rest mass to be zero
   in Dirac equation. Then this transform \cite{lan} in a couple
   of Weyl equations given by the hamiltonians
      $ H _{\pm} = \pm  c \  \mbox{\boldmath  $\sigma \cdot  \hat p$}$
    applied to a two component spinor.
   As usual {\boldmath $\sigma$} denotes 
   the Pauli matrix in vectorial form. Considering $H_-$ and
   assuming that the spin of each particle in the one dimensional
   system is antiparallel to the $x$ axis   
  (i$.$e.\  by considering only the component
   with  $S_x = -1/2$ for  all particle spinors) we obtain 
   the kinetic part of $H_e$, with $\alpha= c$.
  The free waves with positive momentum represent
  a massless particle with spin pointing in the
   opposite direction (a neutrino).
  The free waves with negative momentum, on the other hand, have
  negative energy and opposite helicity
  and should be referred to as
  antiparticles. The problem now is
  the other term in our environment hamiltonian, as 
  it is not easy to imagine an interaction
   depending on positions, which in any case
   should be given in a relativistically correct form
   for particles of this type.
   We may also consider 
   the  electrons in the periodic field of
  crystal atoms, which may be thought of as free
   "dressed electrons". We know\cite{hau} that for an infinite
   simple cubic lattice the one dimensional motion energy curves 
  of these free "quasi particles"  can be 
    described with a good degree of accuracy in the form:
    \begin{displaymath} 
      W(q)_n = A_n - B_n cos(aq) 
    \end{displaymath} 
   where $a$ is the lattice spacing and $q$ the reduced
   wave vector. Interest is usually focused on the neighbours
  at the top and bottom of the band. But for our purposes
  we take one band (for example the first) and develop
  the dispersion relation above near $qa= \pi/2 $,
  assuming the zero of energy scale to be $A_0$:
    \begin{displaymath} 
     W(q)_0 - A_0  = B_0 (aq - \pi/2) = B_0 a q' = E(q') 
    \end{displaymath} 
   with the limitation $|aq'| < 1$.
   We then follow Wannier's method \cite{wan}
   and replace $ q'$ with $ -id/dx$ in the last 
   expression to obtain the effective hamiltonian:
    \begin{displaymath} 
       H = E(-id/dx) = (B_0 a /\hbar) \hat p   \quad .
    \end{displaymath} 
   Our model can therefore be thought of as describing
   a one dimensional system of interacting  quasi
   particles  in a limited pseudo-momentum zone.
 \par
 For the interaction of the environment with
 body $B$ we will assume the linear
 coupling $W^{(sc)} = k X \sum x_i$,
 where $k$ is a constant, already considered
 in the literature. It is similar, for example,
 to von Neumann's measurement interaction \cite{von}.
 This coupling is debatable in that it is not invariant 
  under translation\cite{ford}.
  For our purposes we will use it in this form, but will
  then show that it can safely 
  be replaced by a harmonic coupling.
 If we assume $B$ is free (apart from interaction with the
 environment) and use the approximation mentioned above 
 for a very large $M$, we can write
 the total Hamiltonian as $H = H_e + W^{(sc)} $.
 We will let
    \begin{displaymath} 
  \chi_0   =\psi(X)  \varphi(x_1, x_2, \ldots) 
   \end{displaymath} 
  be the initial state.
 We are interested in its time evolution  and since the commutator
 $ [H_e,W^{(sc)}] $ of the
two terms appearing in the total Hamiltonian commutes with both,
 we can write\cite{mes}:
       \begin{equation}\label{e2}
      \chi_t = e^{-it H_e/\hbar} e^{-it k X \sum x_i/\hbar}
       e^{in\alpha  k X t^2/\hbar^2} \psi(X) \varphi(x_1, x_2, \ldots)     
      \quad .
       \end{equation}
  We now replace  $W^{(sc)}$ with the coupling 
  $W^{(mc)}=\gamma \hat P \sum x_i$, where $\hat P$ is the 
  momentum operator of $B$ and $\gamma$ is a constant.
  By following a similar path it
  is easy to write $ \chi_t$ on momentum basis
  of the macroscopic body:
       \begin{equation}\label{e3}
   \chi_t =\widetilde \psi(P) e^{in\alpha \gamma P t^2/\hbar^2} 
 e^{-itH_e/\hbar}  e^{-it \gamma P \sum x_i/\hbar}
   \varphi(x_1, x_2, \ldots, x_n)
       \end{equation}
  where  $  \widetilde \psi(P) $ is the Fourier transform of 
  $\psi(X)$. In the case of coupling $W^{(sc)}$ we will also consider
  the generic entangled initial state $ \Phi(X, x_1, x_2, \ldots, x_n)$. 
   In this case:
       \begin{equation}\label{e4}
   \chi_t = e^{in\alpha k X t^2/\hbar^2} e^{-itH_e/\hbar} 
  e^{-it kX \sum x_i/\hbar} \Phi(X, x_1, x_2, \ldots, x_n) \quad .
       \end{equation}
 We will use these equations later.
 \vskip .9 true cm
 \noindent{\bf 3. THE REDUCED DENSITY MATRIX }
 \vskip .6 true cm
 \par
 We will first consider the case of coupling  $W^{(sc)}$
  and calculate the macroscopic body's reduced
 density matrix elements $\rho^{(sc)}_{X'X''}$
  on the position basis.
 They are obtained by taking a partial trace of the total
  density matrix, i. e.  integrating $ <X'|\chi_t> <\chi_t|X''>$
  over all the degrees of freedom of the environment:
       \begin{eqnarray}\label{e5}
 &\rho^{(sc)}_{X'X''}&= \psi(X') \psi^{\ast}(X'') e^{i n \alpha k 
 (X'-X'')t^2/\hbar^2} \cdot \nonumber \\  &
      &\int dx_1 \cdots dx_n \varphi^{\ast}(x_1, \ldots, x_n) 
 \varphi(x_1,\ldots, x_n) e^{itk (X'-X'') \sum x_i/\hbar} \ .
       \end{eqnarray}
  We used Eq.~(\ref{e2}) and the
 fact that operator $A =e^{itH_e/\hbar}$ is unitary.

The above equation can be further manipulated by introducing,
 in lieu of $x_i$ the co-ordinate  $ \eta=\sum x_i/n $
 of the centre of mass of the particle system constituting
the environment and the co-ordinates
 of the particles with respect to
their centre of mass:    $ x_i = \eta + \xi_i$.
       The $\xi_i$  are not independent as they must satisfy
       the relation $\sum_{i=1}^n \xi_i = 0$.
       We will consider $\eta$ and 
       the first $n-1$  relative co-ordinates 
 (i$.$e.\ $\xi_1, \xi_2, \cdots, \xi_{n-1}$) as independent variables.
       The last relative co-ordinate
       is expressed as $\xi_n = -\sum_{i=1}^{n-1} \xi_i$.
       The Jacobian determinant of the transformation is equal to $n$.
       We will write the integration volume element for the new
       variables as $dV = d\eta \, dS $, where
       $dS = n d\xi_1 \, d\xi_2 \, \cdots \,  d\xi_{n-1}$.
       By integrating  $  \varphi^{\ast}(\eta + \xi_1, \ldots, ) 
       \varphi(\eta + \xi_1,\ldots, ) $ as expressed in the
       new variables on $dS$  and keeping  $\eta$ fixed, we obtain 
       the quantity 
       \begin{equation}\label{e6}
      w(\eta) = \int dS \varphi^{\ast}(\eta + \xi_1, \ldots, ) 
       \varphi(\eta + \xi_1,\ldots, )
       \end{equation}
    which is the probability density distribution of  $\eta$ in
    the initial state.
    If we define  $z=nk(X'-X'') t/\hbar$ we obtain
       \begin{equation}\label{e7}
 \rho^{(sc)}_{X'X''}=\psi(X') \psi^{\ast}(X'') e^{i n \alpha k 
 (X'-X'')t^2/\hbar^2} \int d\eta e^{iz\eta} w(\eta) \quad .
       \end{equation}
 The Fourier transform $f(z) = FT[w(\eta)] = \int d\eta e^{iz\eta} w(\eta)$
  therefore appears as a factor in the expression of  $\rho^{(sc)}_{X'X''} $.
  In passing we note that at each successive instant, as is apparent
  from the previous expression, the absolute values of the off
  diagonal matrix elements can
  never be bigger than their initial values.
  If we consider the mean value $\overline{\eta}$ of $\eta$, 
 and its mean square deviation 
 $\sigma=\overline{(\eta-\overline{\eta})^2} $ 
  (which we take as finite), for short periods of time we obtain:
       \begin{equation}\label{e8}
       |f(z)|= (1 -\sigma z^2/2)  \qquad .
       \end{equation}
  Now, as $w(\eta)$  is a positive definite quantity with
  mean square deviation $ \sigma \geq 0$,
  in the initial moments the absolute value of the off diagonal
  matrix elements decreases, apart from
 the exceptional case in which  $\sigma=0$.
 But the really interesting point is the asymptotic time behaviour.
  Here we see not only that  $w(\eta)$ is a real
  definite positive function, but also that
 its integral extended from minus to plus infinity must be
 one. So when we have an ordinary function, it is absolutely
 integrable,  with the result that  $|f(z)| \rightarrow 0$ for
  $z \rightarrow \pm \infty$.
 This brings us to the important conclusion that
 starting from any factorized initial state, the off diagonal
 matrix elements $\rho^{(sc)}_{X'X''}$ (excluding exceptional cases)
 go asymptotically to zero in the two
 time directions. To this end we should remember that if  $w(\eta)$
  is  $m$ times continuously differentiable and its $m$ derivatives
  are integrable, we obtain  $|f(z)| \rightarrow 0$ for
  $z\rightarrow \pm \infty$ more quickly than $1/z^m$.
  If $w(\eta)$ is one of the more commonly encountered
  probability density distributions
  (such as a Lorentz or a Gaussian distribution),
   the off diagonal matrix elements tend to zero like
 the related Fourier transform, an exponential or a
 Gaussian, respectively. When the environment initial state vector 
 is factorized in the states of the individual components, the
 probability distribution  $w(\eta)$ is a Gaussian\cite{car} under very 
 general and physically acceptable conditions.
 \par 
 Additional insights into the result obtained above are provided
  by the following examples, 
  which consider the unfavourable case of a non
 differentiable function such as
       \begin{displaymath} 
   w(\eta) = \cases { 1/2L & if $|\eta| < L$  \cr 0 & otherwise \cr}
       \end{displaymath} 
 and the case in which  $w(\eta)$  is a non-ordinary and somewhat
 singular "function" such as a  $\delta(\eta-\overline \eta)$.
  In the first case we have
 \begin{equation}\label{e9}
 f(z)=\sin [z L]/zL = \sin [nk(X'-X'') L t /\hbar]/[nk(X'-X'') L t /\hbar]
 \end{equation}
 which allows us to define a decoherence time scale
 $ \tau^{(sc)} = \hbar/[n k (X'-X'') \Delta \, \eta]$, where
   $\Delta \, \eta$ is the width of  $w(\eta)$.
  In the second we have
 \begin{displaymath}
  f(z)=e^{i z \overline \eta} \qquad .
 \end{displaymath}
 As we can see, the absolute value
  $\rho^{(sc)}_{X'X''}$ is now a constant of motion but in
 spite of this the phase oscillations, which get increasingly faster, might
 help towards practically cancelling out the spatial correlations.
 So far we have assumed that environment wave
 function $\varphi(x_1, x_2, \ldots, x_n)$ is normalizable
  in the usual sense (especially with respect 
 to the variable $\eta$),
 i.e. as a square integrable wave function.
  But we also have to consider the case in which, for instance, 
  the state of $\eta$ is that of a pure plane wave.
  In these circumstances we could resort to box normalization 
  and obtain our result simply by taking the limit for $L$
  going to infinity in  Eq.~(\ref {e9}).
  In this case decoherence is immediate.
 \par
  We will now consider the case of linear
  coupling $W^{(mc)}$ rather than coupling $W^{(sc)}$.
  Using  Eq~(\ref{e3}) and following, mutatis mutanda,
  the steps we took before, we
 now see that the off diagonal elements  $ \rho^{(mc)}_{P'P''}$
  of the macroscopic body's
 reduced density matrix on the momentum basis 
  depends on time as:
       \begin{equation}\label{e10}
 \rho^{(mc)}_{P'P''}=\widetilde \psi(P')
 \widetilde \psi^{\ast}(P'') e^{i n \alpha \gamma 
 (P'-P'')t^2/\hbar^2} \int d\eta e^{iy\eta} w(\eta)
     \end{equation}
       where $y=n\gamma(P'-P'') t/\hbar$.
       The value of $|\rho^{(mc)}|_{P'P''}$
  (as is the case for $W^{(sc)}$ interaction) depends
       on the Fourier transform $f(y) = FT [w(\eta)]$,
       which means that time behaviour
       is the same as in the previous case,
       replacing $z$ with $y$.
 \par
       The results we obtained for both 
       couplings have a very plausible interpretation.
       Looking at Eq~(\ref{e2}) and Eq~(\ref{e3}) we see
       that the initial states of the environment,
       coupled respectively to $X'$ and $X''$  (Eq~(\ref{e2}))
       or to $P'$ and $P''$ (Eq~(\ref{e3})), simply " move apart"
       in Hilbert space (actually in both directions of time).
       Indeed 
       \begin{displaymath}
         e^{it k X' \sum x_i/\hbar} \ \varphi(x_1, x_2, \ldots)     
       \end{displaymath}
       gives a momentum translation amounting to $ tkX'$ for each
       particle of the environment, i$.$e.\ to 
      $tkX'n$ for their centre of mass (c.m.). The two
       wave packets of the c.m. (coupled to $X'$ and $X''$ or
       to $P'$ and $P''$ respectively)
       then become separate on a momentum basis
       by the quantity $tk|X'-X''|n$
       (or $t\gamma|P'-P''|n$) and at some time
       no more overlap. If the mean square deviation
       of the environment's c.m. momentum
       is denoted as $\Delta \pi$, we may suppose that
       decoherence is established after
       a time $\tau^{(sc)} = \Delta \pi/kn|X'-X''|$
       (or $\tau^{(mc)} =\Delta \pi/\gamma n|P'-P''|$ respectively).
       We may use the indeterminacy relation to write
       $\tau^{(sc)} = \hbar/kn|X'-X''|\Delta \eta$, thus
       recovering the definition already given of
       a "decoherence time" scale.
  \par
  Let's go back to the spatial coupling  $W^{(sc)}$ and take
  a generic wave function  $\Phi(X, x_1, x_2, \ldots, x_n)$ as the
  initial state. We assume it is square integrable and
  continuous with continuous first derivatives for all variables.
  Using  Eq~(\ref{e4}) we obtain:
       \begin{equation}\label{e11}
        \rho^{(sc)}_{X'X''}= e^{i n \alpha k (X'-X'')t^2/\hbar^2} g(z)
       \end{equation}
  where  $g(z) = FT[w_{X'X''}(\eta)]$, with an obvious extension of
  the notation, is the Fourier transform of:
       \begin{equation}\label{e12}
 w_{X'X''}(\eta)=\int \Phi(X', \eta+\xi_1, \ldots, \eta-\sum_1^{n-1} \xi_i)
  \Phi^{\ast}(X'', \eta+\xi_1, \ldots, \eta-\sum_1^{n-1} \xi_i)  dS 
       \end{equation}
 The co-ordinates transformation thus introduced does not affect the
 assumed square integrability or the analytical properties of the initial state
 wave function expressed in the new variables. Evidently 
   $w_{X'X'}(\eta)=|w_{X'X'}(\eta)|$ and
       $w_{X''X''}(\eta)=|w_{X''X''}(\eta)|$ are
 absolutely integrable in $\eta$, and therefore their square root
  will be square integrable. If we apply Schwarz  inequality to
  the integration in $dS$, we obtain
 \begin{equation}\label{e13}
       |w_{X'X''}(\eta)| \leq w_{X'X'}^{1/2}(\eta) w_{X''X''}^{1/2}(\eta)
   \qquad .
 \end{equation}
 The right hand quantity is integrable in $\eta$, as is clear
  if we again apply
 Schwarz inequality and take the properties of the two
 functions involved into account. The limitation in  Eq.~(\ref{e13})
  implies that  $w_{X'X''}(\eta)$
 is absolutely integrable in $\eta$ and, for the
  properties of the Fourier transform already used,
  the off diagonal matrix elements starting 
 from the initial entangled state  $\rho^{(sc)}_{X'X''}$ tend to zero
  for  $z \rightarrow \pm \infty$, i.e. in the two time directions.
 \par
 Lastly we will consider the harmonic coupling
       \begin{displaymath}
       W^{(hc)} = -1/2 k \sum_i(X-x_i)^2 = -1/2 kn X^2 + k X \sum x_i
       -1/2 \sum x_i^2  \quad .
       \end{displaymath}
 The last term is absorbed in  $V(x_1,x_2, \ldots)$, the first
  only contributes a phase
 factor to the matrix elements  $\rho^{(hc)}_{X'X''}$
  and the rest is the linear coupling we have
 already seen. Starting from the same initial conditions as in the
 case of $W^{(sc)}$, we obtain
 \begin{equation}\label{ee10}
 \rho^{(hc)}_{X'X''}=\rho^{(sc)}_{X'X''} e^{-i k n (X'^2 -X''^2) t/[2 \hbar]}
    \quad .
 \end{equation}
  For the sake of completeness we also consider the analogous 
  quadratic coupling :
       \begin{displaymath}
       W^{(mhc)} = -1/2 \mu \sum_i(P-\nu x_i)^2 = -1/(2 \mu)n P^2
      + \gamma P \sum x_i -\gamma \nu/2 \sum x_i^2  \ .
       \end{displaymath}
  with $\gamma=\nu/\mu$.
 Here too the last term is absorbed in  the potential
  part of the environment hamiltonian, the first
  only contributes a phase
  factor and the rest is just the coupling $W^{(mc)}$.
   Starting from the initial (factorized) state
   considered in that case:
 \begin{equation}\label{ee11}
 \rho^{(mhc)}_{P'P''}=\rho^{(mc)}_{P'P''} e^{-i n (P'^2 -P''^2) t/[2 \mu \hbar]}
  \qquad .
 \end{equation}
\vskip .9 true cm
 \noindent{ \bf 4. CONCLUDING REMARKS }
 \vskip .6 true cm
 \par
 The most significant result obtained using this environment model, 
  in the limit of infinite mass for the macroscopic body,
 has been to show that decoherence occurs on the position basis
 in the case of linear coupling $W^{(sc)}$ and harmonic coupling
 under very general initial conditions. We have only proven that
 decoherence occurs as time is going towards infinity,
 and we cannot therefore exclude the hypothesis that the spatial
 correlations of body $B$
 may increase during finite periods.
 However, when the initial state was factorized we saw that the absolute
 values of the off diagonal elements of the reduced density matrix not only
 go asymptotically to zero but also can never exceed their 
 initial value. That particular initial condition so look
 as preferred with respect this point.
 In the case of this type of initial condition we have also showed the 
 time dependency with which the matrix elements in question go
 asymptotically to zero. The speed at which they do so depends
 on the initial state of the environment.
 \par
   In the case of coupling  $W^{(sc)}$ we defined a
   time scale $ \tau^{(sc)} = \hbar/[n k (X'-X'') \Delta \, \eta]$, where
   $\Delta \, \eta$ is the 
  initial distribution width of the environment's centre
   of mass position.
   We may assume that decoherence is established when
   $|t|  \gg \tau^{(sc)}$. We have not attempted a quantitative
   estimate of $\tau^{(sc)}$  as our model is far
  from realistic. Qualitatively,
 however, we observed that as well as depending on the environment
 through the initial mean square deviation of  $\eta$, it is inversely
  proportional to the
 difference $X' - X''$, to the strength of coupling constant and to
 the number of particles $n$ with which the macroscopic body interacts.
  This sort of dependence is certainly reasonable and in passing we note that
 decoherence occurs, of course after a greater time interval,
  even with very small coupling constants and through interaction
  with a single particle. It follows that the influence of the
  environment on the quantum
 state of a macroscopic body cannot be ignored and has dramatic effects,
 however weak.
 \par
  Still in the case of a factorized initial state, we saw how the preferred
  basis (i.e. on which the macroscopic body's reduced 
  density matrix becomes diagonal)
 depends on the type of interaction with the environment. It is
  the position basis in the case of linear $W^{(sc)}$ and harmonic
 coupling and the momentum basis in the case of  $W^{(mc)}$ coupling.
  The preferred basis
  in our case is the one of the operator commuting with the interaction
 Hamiltonian. The importance of an interaction commuting
 with the preferred basis was first stressed by
  Zurek\cite{zur}. Our results so provide further
 confirmation of what is generally taken for granted,
  i.e. that the natural preferred
 basis for a macroscopic body is the position basis, because the system
 interacts with the rest of the world through spatial co-ordinates.
 \par
 Another point to note is that our results are independent of whether or not
 there is an interaction potential between the particles making up the
 environment. This means, in our example at least, that 
 decoherence need not be thought of as linked to a thermalization process. 
 On the other hand,
 this could also be inferred from the case of decoherence due to
  scattering\cite{zeh,gal,teg}
 and, in fact, our model describe
  a scattering by independent particles when there 
 is no interaction between them.
 \par
  To conclude we note that in the present model, which
 assume the whole system of macroscopic body plus the environment
  as a closed system, decoherence
 is not an irreversible process\cite{kie}. This is 
 made clear in our study by following the evolution of
   the pure state of the whole
  system, and because the results obtained are asymptotically
  valid in the two time directions.
 The situation would be quite different for a more realistic
 permanent flux of incoming particles without initial
 correlation.
  But our model allows for processes with suitable
 initial conditions (though difficult to realize in practice
 and which in our case cannot be factorized) for which the
 spatial correlations of the
 macroscopic body increase over a finite time.
  None of this prevents decoherence from
 occurring asymptotically. In my opinion this situation calls to mind the
 spread of the wave packet of a free particle. Depending on the initial
 conditions it may even contract at the outset, but it will always end up
 expanding asymptotically.
       \vskip 1.4 true cm

      \end{document}